\documentclass[conference]{IEEEtran}
\IEEEoverridecommandlockouts
\usepackage{setspace}
\usepackage{cite}
\ifCLASSINFOpdf
\usepackage[pdftex]{graphicx}
\else
\usepackage[dvips]{graphicx}
\fi
\usepackage{amsmath,amssymb,amsthm}
\usepackage{array}
\usepackage{algorithm,algorithmic}
\usepackage{subfigure}
\usepackage{makecell}				% bold horizontal table line
\usepackage{cite}
\usepackage{multicol}
\usepackage{color}
\ifCLASSOPTIONcompsoc
\usepackage[caption=false,font=normalsize,labelfont=sf,textfont=sf]{subfig}
\else
\usepackage[caption=false,font=footnotesize]{subfig}
\fi
\newcommand{\PreserveBackslash}[1]{\let\temp=\\#1\let\\=\temp}
\newcolumntype{C}[1]{>{\PreserveBackslash\centering}p{#1}}
\newcolumntype{R}[1]{>{\PreserveBackslash\raggedleft}p{#1}}
\newcolumntype{L}[1]{>{\PreserveBackslash\raggedright}p{#1}}

\usepackage{geometry}
\begin{document}
% \title{Multi-dimensional Network Resource Integration architecture Towards Metaverse-as-a-Service in 6G}
\title{Slicing4Meta: An Intelligent Integration Architecture with Multi-dimensional Network Resources for Metaverse-as-a-Service in Web 3.0}
\author{
  Yi-Jing Liu, Hongyang Du, Dusit Niyato,~\IEEEmembership{Fellow,~IEEE}, Gang Feng, Jiawen Kang,
  Zehui Xiong 
% }%
\thanks{©2023 IEEE. Personal use of this material is permitted. Permission
from IEEE must be obtained for all other uses, in any current or future
media, including reprinting/republishing this material for advertising or promotional purposes, creating new collective works, for resale or redistribution to servers or lists, or reuse of any copyrighted component of this work in other works}
}

\maketitle

\begin{abstract}
 As the next-generation Internet paradigm, Web 3.0 encapsulates the expectations of immersive and interactive experiences in a decentralized manner. Metaverse, a virtual world interacting with the physical world, is becoming one of the most potential applications to push forward with Web 3.0. In the Metaverse, users expect to tailor immersive and interactive experiences by customizing real-time Metaverse services (\emph{e.g.}, augmented/virtual reality and digital twin) in the three-dimensional virtual world. Nevertheless, there are still no unified solutions for the Metaverse services in terms of orchestration and management. Therefore, continuous and seamless upgrading of network systems is essential yet challenging to support Metaverse services. In this paper, to provide scalable solutions for tailoring Metaverse services, we propose a new concept, named Metaverse-as-a-Service (MaaS), where users only need to purchase the MaaS models that are needed (\emph{e.g.}, computing, communications, and data resources) to achieve their required Metaverse services. Furthermore, to unify the orchestration and management of MaaS models, we propose a novel architecture, called Slicing4Meta, to customize Metaverse services by integrating MaaS models and the associated multi-dimensional resources. Additionally, we propose the classification for typical Metaverse services based on the quality of experience (QoE) requirements and illustrate how to fulfill the QoE requirements under the Slicing4Meta architecture. We then illustrate a virtual travel study case, in which we examine the relationship between the QoE and the multi-dimensional resources by quantitatively modeling the QoE of Metaverse users. Finally, we discuss some open challenges of Slicing4Meta and propose potential solutions to address the challenges.

 % It is calling for a continuous and seamless upgrading of network systems to support Metaverse services, and thus bring Metaverse into reality. 

\end{abstract}

\IEEEpeerreviewmaketitle

\section{Introduction}
Web 3.0 is defined as the next generation World Wide Web, which revolves around user-centric services that require immersive, interactive and simulating experiences. By promising to support these services, Metaverse is becoming one of the most potential applications to envision Web 3.0. The core idea of the Metaverse is to build a virtual world that can interact with the physical world, while providing continual service experiences of real-time social interaction between consumers, creators and businesses. However, Metaverse services which are a collection term for services supported in Metaverse, such as augmented reality (AR)/virtual reality (VR), and 3D holographic communications, cannot be supported by the current networks well. On the one hand, the quality of experience (QoE) requirements of Metaverse services, such as continuous connection availability (99.99999\% reliability) and ultra-low interaction time (sub-millisecond) for 3D holographic display \cite{8869705,speculative,AI-assisted}, may go beyond the capability of current networks. On the other hand, there are no unified solutions for the current network systems to flexibly orchestrate and manage Metaverse services. Therefore, continuous and seamless upgrading of network systems is essential yet challenging to support Metaverse services.

The network systems for supporting Metaverse services have yet to be developed. However, it is foreseeable that the eventual network system will be a convergence of a service-oriented pespective and an evolutionary perspective. Specifically, the service-oriented perspective focuses on meeting diverse QoE requirements over a common network infrastructure. This requires to tailor and isolate the specific services on-demand, as the services should independently operate without any influence \cite{Minrui,9040264}. The evolutionary perspective is adopted to scale up and improve network capabilities significantly, which requires a more flexible and scalable architecture to orchestrate and manage Metaverse services.

On the one hand, ``as-a-service (aaS)'' can provide a flexible solution for tailoring and isolating specific Metaverse services. In the last decade, with the arrival of software-as-a-service providers, platform-as-a-service and infrastructure-as-a-service solutions, the rise of everything-as-a-service (XaaS) has been phenomenal. Indeed, XaaS means that everything over cloud systems can be regarded as a service, which is a business model that converts the customer-supplier relationship from the traditional ownership model to a new model providing services on a non-ownership basis \cite{banerjee2011everything}. Intuitively, Metaverse can benefit from aaS models. In Metaverse,
% MaaS is a collection of aaS models in Metaverse, which is defined as an on-demand subscription solution that allows individuals/organizations to develop and strengthen various forms (\emph{e.g.}, presence, management, and implementation) in Metaverse to support business processes, collaboration, immersive and real-time interaction and other related use cases. 
% Expanding from aaS models in cloud systems, everything (\emph{e.g.}, resources, information and communication technology (ICT) functions, technologies and data) in Metaverse can be delivered as a service
we propose Metaverse-as-a-Service (MaaS), which is a collection of physical or virtual business models. Specifically, users only need to pay for the MaaS models (\emph{e.g.}, computing, communications, and data resources) needed to achieve their required Metaverse services, instead of traditional purchases or license models that require fixed, upfront payment regardless of usage volume. By offering various consumption and payment schemes for users, service providers can deliver an alternative solution consisting of MaaS models to users.

On the other hand, some prior studies have suggested that network systems should be constantly upgraded by integrating multi-dimensional heterogeneous resources such as communication resources, computing resources, storage resources, and even various radio access technologies (RATs) to scale up and improve network capabilities \cite{8869705,speculative}. Indeed, all MaaS models are built upon various multi-dimensional network resources (Part II in Fig. 1) in the form of supplying and/or consuming certain resources.
% As shown in Fig. 1 (Part I), MaaS mainly consists of
% component-as-a-service (CaaS) that supplies multi-dimensional resources and technology-as-a-service (TaaS) that consumes multi-dimensional resources. 
% By packaging, integrating, configuring and deploying multi-dimensional resources, orchestration and management entities (\emph{e.g.}, controllers) are expected to modify, create and deliver MaaS models to MSIs, while improving the network capabilities.
By integrating these MaaS models and multi-dimensional heterogeneous resources, orchestration and management entities (\emph{e.g.}, controllers) are expected to modify, create and deliver MaaS models to MSIs, while making the network capabilities improvement possible.

% On the other hand, some prior studies have suggested that network systems should be constantly upgraded by integrating multi-dimensional heterogeneous resources such as communication resources, computing resources, storage resources, and even various radio access technologies (RATs) to scale up and improve network capabilities \cite{8869705,speculative,9382026}. By integrating MaaS models and the relevant multi-dimensional heterogeneous resources, network systems can flexibly construct MSIs by modifying existing MaaS models or creating new MaaS models, thus achieving cost-efficient and performance-optimal orchestration and management of network systems. }

\begin{figure}[htbp]
\centering
\vspace{-2mm}
\includegraphics[width=8cm]{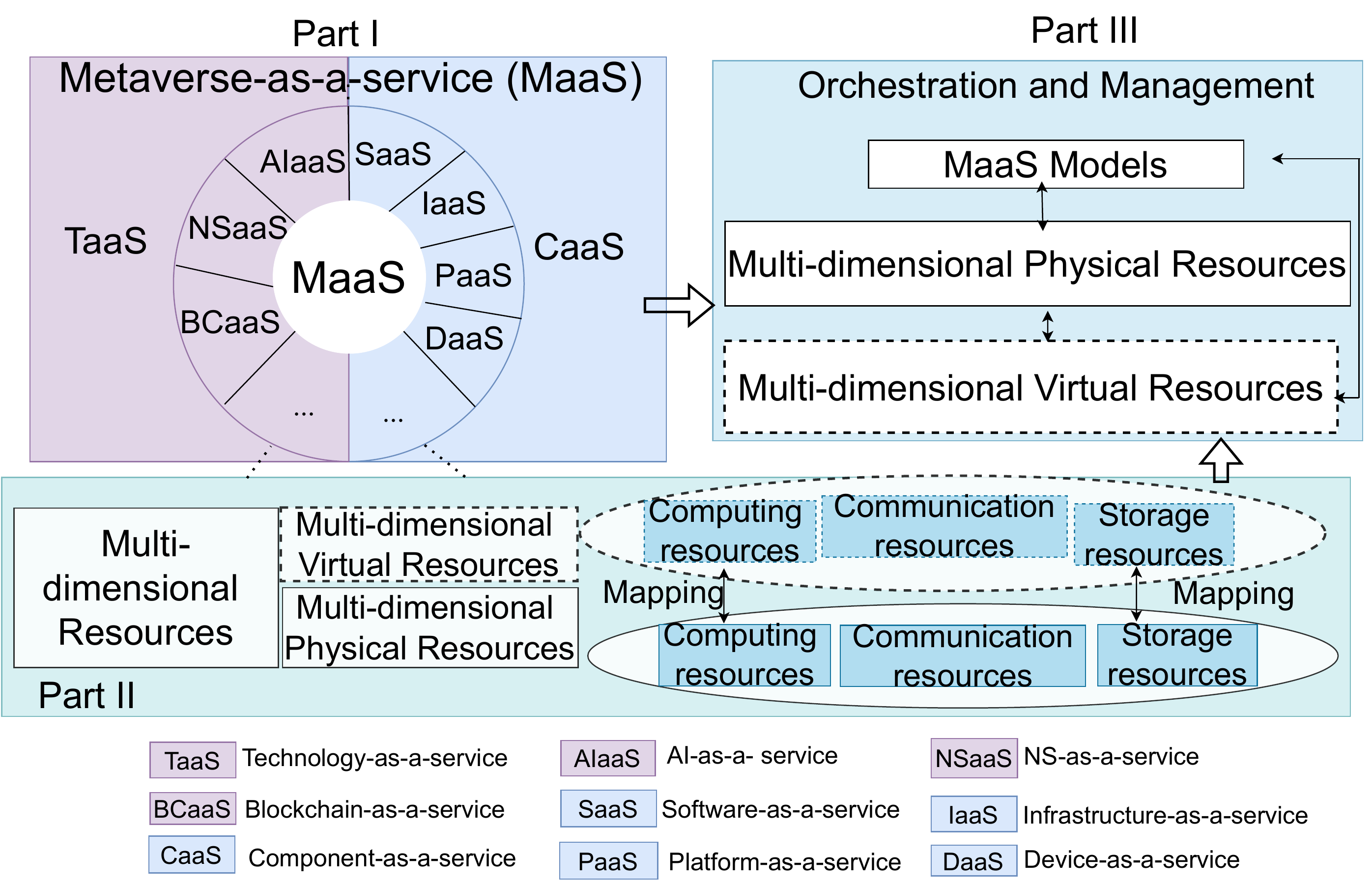} 
\caption{Metaverse-as-a-service: everything in Metaverse can be delivered as a service}
\label{fig system model}
\vspace{-2mm}
\end{figure}

However, it is rather complicated for network systems to integrate various MaaS models for diverse Metaverse services, especially when configuring and managing resources, functions, technologies and data for MaaS models. This is due to the dynamic nature of the Metaverse environments and the high computation complexity incurred by integrating MaaS models and multi-dimensional resources. To cope with the aforementioned problems, artificial intelligence (AI) could be an efficient way to configure and manage networks in an intelligent and autonomous fashion \cite{AI-assisted,guan2021customize}. For example, the AI-assisted architecture can timely predict and update physical-virtual world status and service requirements, from which network operators can deploy or configure the needed MaaS models in advance [9].

In this paper, we propose an evolved network system architecture for providing customized Metaverse services, named {\bf{Slicing4Meta}}, by introducing intelligent integration of MaaS models and multi-dimensional resources. The main contributions of this paper are summarized as follows:

\begin{enumerate}
\def\labelenumi{(\arabic{enumi})}
\item
To tailor subscription solutions for Metaverse services while fulfilling diverse QoE requirements requested by these services, we propose a new concept, namely Metaverse-as-a-Service (MaaS), where various components and technologies in Metaverse can be delivered as services.

% We propose MaaS to facilitate the holistic construction of on-demand subscription solutions by delivering various components and technologies as services. Based on MaaS, we propose Slicing4Meta architecture to integrate MaaS models and the associated multi-dimensional resources and thus customize Metaverse services to meet the specific QoE requirements.

% integrating multi-dimensional resources from various components and technologies.

% intelligently integrate multi-dimensional resources under the HyperSlicing architecture to support Metaverse services by introducing a variety of advanced communication, networking, information, and intelligent technologies.

% \item
% We propose HyperSlicing architecture to integrate MaaS models and the associated multi-dimensional resources and thus customize Metaverse services to meet the specific QoE requirements.
% and thus achieve the cost-efficient and performance-optimal orchestration and management.

\item
To unify the orchestration and management of MaaS, we propose a novel intelligent resource integration architecture, called Slicing4Meta.
% , which introduces two-tier AI-assisted controllers to facilitate integrating MaaS models and the associated multi-dimensional resources. 
Different from the conventional types of network services, we define two typical types of Metaverse services, including VR/AR immersive services 
% (\emph{e.g.}, game and virtual travel)
 and digital twin services 
 % (\emph{e.g.}, smart home and manufacturing), 
 to facilitate the holistic construction of Metaverse service instances. 

% We define two typical types of Metaverse services based on the QoE requirements, including VR/AR services (\emph{e.g.}, game and virtual travel) and digital twin services (\emph{e.g.}, smart home and manufacturing). 

% a variety of advanced communication, networking, and information technologies are introduced to assist in 

\item
We illustrate a virtual travel case, where we quantitatively examine the relationship between the QoE of users and multi-dimensional resources under the Slicing4Meta architecture. Additionally, some open challenges including isolation and security under Slicing4Meta and the potential solutions are presented.

% \item
% We evaluate the performance gain of the specific Metaverse scenarios under HyperSlicing.
\end{enumerate}

% \begin{figure*}[htbp]
% \centering
% \vspace{-8mm}
% \includegraphics[width=12cm]{Fig2.pdf} 
% \caption{Slicing4Meta architecture: an intelligent integration architecture with multi-dimensional network resources for Metaverse-as-a-Service}
% \label{fig system model}
% \vspace{-5mm}
% \end{figure*}

\section{Slicing4Meta for Metaverse Services}\label{sec:model}
In this section, we first give the details of MaaS. Then we propose an evolved network architecture for network systems, called Slicing4Meta, to support Metaverse services by customizing various MaaS models. Finally, we present the details for the instantiation of Metaverse services.

\subsection{Metaverse-as-a-service (MaaS)}
% In this section, we first define MaaS according to the needs of Metaverse services and propose two types of MaaS models. Then, we present specific solutions by delivering various MaaS models.

% \subsection{Definition and Types of MaaS}
% MaaS is defined as an on-demand subscription solution that allows businesses and/or operators to develop and enforce various forms ( \emph{e.g.}, presence, management, orchestration and implementation) in Metaverse to support Metaverse service processing, collaboration, business operation, products and other related scenarios. Similar to the XaaS in cloud systems, everything in Metaverse can be regarded as a delivery model, which can be easily created and/or modified as function modules. 

% Building on MaaS, businesses and operators can create virtual identities and events, and collaborate in virtual spaces. 

% could be delivered as a service model

As shown in Fig. 1 (Part I), MaaS consists of component-as-a-service (CaaS) and technology-as-a-service (TaaS). The CaaS mainly refers to the as-a-service models that supply multi-dimensional resources such as software-as-a-service (SaaS).
 % and infrastructure-as-a-service (IaaS). 
% SaaS refers to a software delivery model where software providers host the applications and make them accessible over the Internet. 
 % IaaS is a network infrastructure resource delivery model that provides network resources with scalability, such as servers and base stations. 
The TaaS mainly refers to the as-a-service models that consume multi-dimensional resources such as network slice-as-a-service (NSaaS). 
 % (AIaaS) and network slice-as-a-service (NSaaS). 
 % AIaaS refers to off-the-shelf AI tools that enable businesses/operators to implement, use, and/or scale AI technologies and algorithms. 
% NSaaS refers to the delivery model in terms of functionalities, topology, policies, and parameters that are mapped from Metaverse service demands.

A MaaS model could be a single delivery model or a composite delivery model that consists of multiple individual MaaS models. By packaging, modifying, delivering and integrating MaaS models, businesses and/or operators can quickly provide customized solutions to support Metaverse services. Indeed, there are a large number of alternative ways to form the required MaaS models, which increases the difficulty in instantiating Metaverse services especially when the number of services is large. Therefore, a resource integration architecture is needed to configure optimal MaaS models for specific services while guaranteeing QoE requirements.

\subsection{Slicing4Meta Architecture} 
Currently, many representative institutions have been considering developing software-defined networks (SDN) and network function virtualization (NFV) technologies in future network systems. For example, 3GPP TR 28.801 has announced that network slicing can be enriched by using NFV and SDN in terms of orchestration and management, network function deployment and resource allocation \cite{3GPP2018}. China Mobile Communication is designing a lite SDN and NFV-enabled network for 6G networks \cite{9205980}. Along with future networks, Metaverse is expected to be enabled by SDN and NFV technologies.
\begin{figure*}[htbp]
\centering
\vspace{-8mm}
\includegraphics[width=12cm]{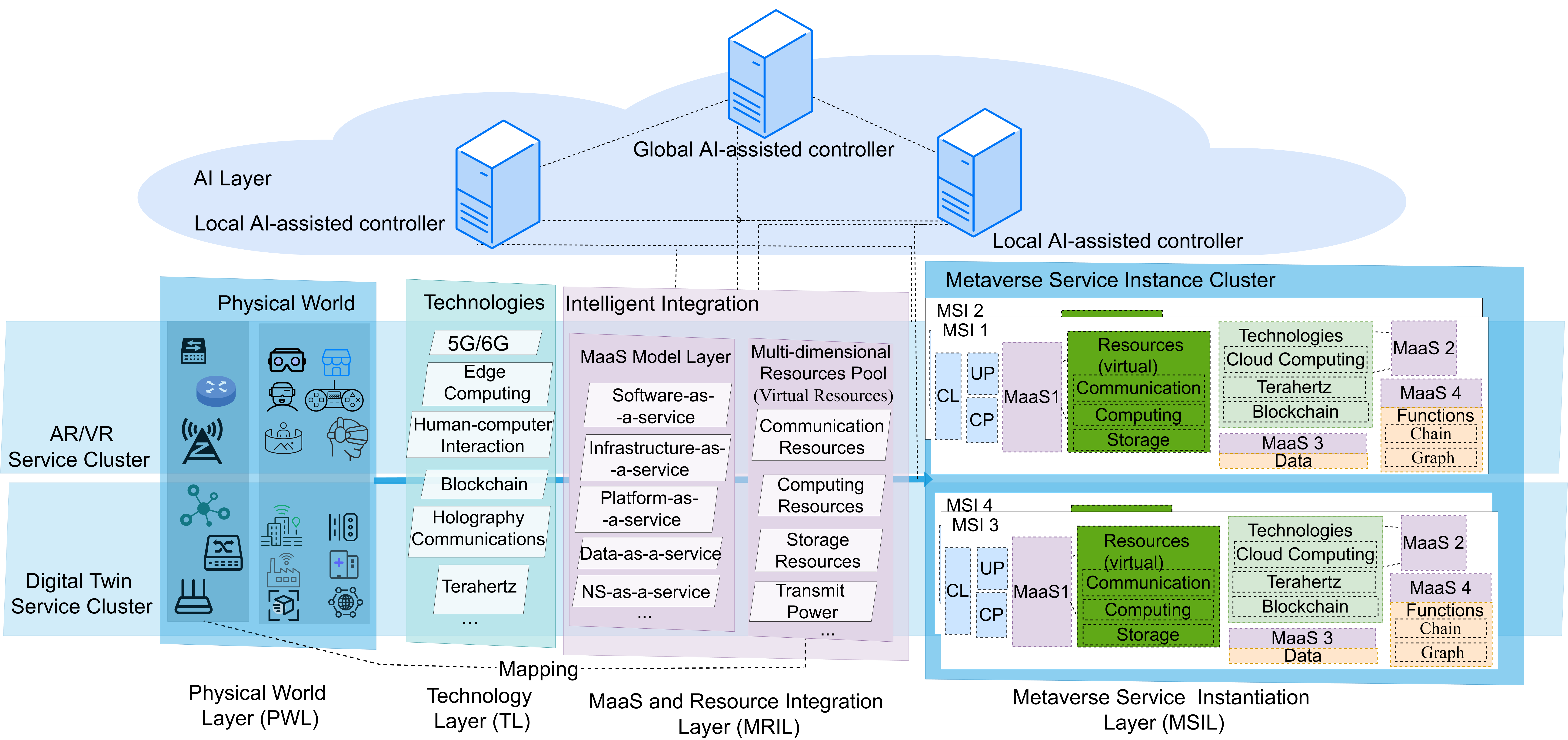} 
\caption{Slicing4Meta architecture: an intelligent integration architecture with multi-dimensional network resources for Metaverse-as-a-Service}
\label{fig system model}
\vspace{-5mm}
\end{figure*}

% Currently, many representative institutions have been considering developing SDN and virtualization technologies in future network systems. For example, 3GPP TR 28.801 has announced that network slices can be enriched by using NFV and SDN in terms of orchestration and management, network function deployment and resource allocation \cite{3GPP2018}. China Mobile Communication is designing a lite SDN and NFV-enabled network for 6G networks \cite{9205980}. Along with future networks, Metaverse is expected to be enabled by software-defined networks (SDN) and virtualization technologies.}
% Researchers from both academia and industry have widely agreed that Metaverse should be enabled by the software-defined network (SDN) and virtualization technologies. As per this agreement, we propose an evolved architecture, namely Slicing4Meta, as shown in Fig. 2.

% controllers to maintain a global view of Metaverse and thus integrate and manage specific MaaS models and the corresponding physical-virtual resources for achieving Metaverse services.

Slicing4Meta consists of five layers, including the physical world layer (PWL), technology layer (TL), MaaS and resource integration layer (MRIL), Metaverse service instance layer (MSIL), and AI layer. Different from the network slicing composed of service instance layer, network slice instance layer and resource layer in 3GPP TR 28.801, we specify two new layers including the TL and AI layer in Slicing4Meta, while PWL, MSIL and MRIL are expanded from network slicing. Specifically, the TL consists of the technologies that are needed during the whole procedures of processing service, such as human-computer interaction used to connect the physical-virtual worlds and Terahertz used to guarantee ultra-high data rates. The MRIL includes MaaS model layer and a virtual multi-dimensional resource pool, which is used to configure the optimal MaaS models with certain resources for specific Metaverse services. Note the virtual resources in resource pool are generally mapped from the physical resources. In addition, the PWL consists of the objects in the real world such as physical network resources, physical devices and physical entities. Depending on the service requirements, we divide the Metaverse services into two clusters: (1) AR/VR service cluster that requires immersive and social interaction experiences, and (2) DT service cluster that focuses on creating high-fidelity virtual models for physical objects to simulate their behaviors. In the MSIL, an Metaverse service instance (MSI) represents an instance of Metaverse service, whereas an MSI cluster refers to a set of service instances with the same service type. The AI layer consists of local AI-assisted controllers and a global AI-assisted controller, where AI technologies are integrated into SDN controllers to intelligently orchestrate and manage various MaaS models. Specifically, the global AI-assisted controller located at a central cloud maintains a global view of Metaverse to integrate and manage MSIs of different types. The local AI-assisted controller located at MSI clusters is to manage/orchestrate MSIs of the same type by creating/modifying MaaS models and the corresponding resources to users.

\subsection{Instantiation of Metaverse Service}
% The incremental deployment (\emph{e.g.}, the functions, technologies, data, procedures and interfaces) should be activated when instantiating metaverse-related services. 
Generally, different Metaverse services are provided by Metaverse service providers through using different MSIs, where different MSIs may share the same MaaS models (\emph{e.g.}, MaaS 2 can be shared by both MSI 1 and MSI 2 in Fig. 3). In addition, some Metaverse services with similar QoE requirements, may be instantiated as the same MSI (\emph{e.g.}, AR service 1 and AR service 2 in Fig. 3) by using an existing MSI or creating a new MSI. In the context of current network slicing, the network service orchestration and management are responsible for managing the lifecycle of network slices instances and orchestrating the infrastructure resources \cite{3GPP2018}. Upgrading from current network slicing, Slicing4Meta's orchestration and management are engaged to manage the lifecycle of MSIs, while orchestrating the infrastructure resources, technologies and data. Indeed, the orchestration and management of Metaverse service are similar to that of current network services, where the difference is that the Slicing4Meta integrates the MaaS model that consists of network functions, technologies, data and relevant resources, instead of only integrating resources. Inspired by network slicing in 3GPP TR 28.801 \cite{3GPP2018}, we propose the following three instantiation-related management functions to manage MSIs and thus support Metaverse services.

% the MSIs are similar to the customized network slice instance . The main differences between them are the heterogeneous resources integration, the associated technologies and the supported scenarios. 

% Metaverse service sub-instances (MSSIs). In the context of current 5G networks, the MSIs/MSSIs are similar to the network slice instance/subnet instance \cite{3GPP2018}. The differences between them lay on the heterogeneous resources integration, the communication/network/information/intelligent technologies, as well as the supported scenarios. 

\subsubsection{Metaverse Service Management Function (MSMF)} As shown in Fig. 3, MSMF is responsible for translating the requirements of users in the physical world to the requirements of Metaverse services in the virtual world via some advanced technologies (\emph{e.g.}, human-computer interaction) and/or devices (\emph{e.g.}, AR/VR glasses). Moreover, it is responsible for managing Metaverse services provided by virtual service providers, where MSMF should be notified about any changes in terms of MSIs, MaaS models, network capability, multi-dimensional resources, and service requirements in physical-virtual worlds.

\subsubsection{Virtual orchestration and management Function (VMOF)} VMOF is responsible for managing and orchestrating MSIs, controlling the life-cycle of MSIs (preparation, planning, run-time and decommission) \cite{3GPP2018}, and interacting with the domain-specific orchestration. Moreover, it can either reuse an existing MSI or create a new MSI from the perspective of the global Metaverse domain by monitoring the performance of MSIs.

\subsubsection{MaaS Management Function (MMF)} MMF is responsible for instantiating, configuring, managing and orchestrating MaaS models and the corresponding resources. Moreover, it analyzes the requests from VMOF and determines the MaaS models and multi-dimensional resources to be used/modified. 

Note when a service arrives, some functions used for instantiating and processing service such as translating the requirements of services and controlling the life-cycle of service instances, are always necessary to achieve the service whatever the service is. The differences in the three management functions between network slicing and Slicing4Meta mainly come with the differences between 5G services and Metaverse services. As shown in Fig. 3, the instantiation process of MSIs consists of three steps, as follows:
\begin{itemize}
\item {\textbf{Step I:}} When users send the Metaverse service requests, the MSMF first translates/updates the \emph{Metaverse services related requirements} (\emph{e.g.}, service type and QoE requirements) and then sends them to VMOF.
\item {\textbf{Step II:}} Once VMOF receives the \emph{Metaverse services related requirements}, it converts them to \emph{Metaverse service sub-instance related requirements} (\emph{e.g.}, MaaS models, multi-dimensional resources and QoE requirements) and then sends the requirements to MMF.
\item {\textbf{Step III:}} The MMF analyzes the requirements from VMOF. Moreover, MMF and VMOF communicate with each other to decide whether to modify existing MSIs and MaaS models and/or create new MSIs and MaaS models.
\end{itemize}

% Step \uppercase\expandafter{\romannumeral1}:  Step  \uppercase\expandafter{\romannumeral2}: . Step \uppercase\expandafter{\romannumeral3}: 

\begin{figure}[htbp]
\centering
\vspace*{-4mm}
\includegraphics[width=8cm]{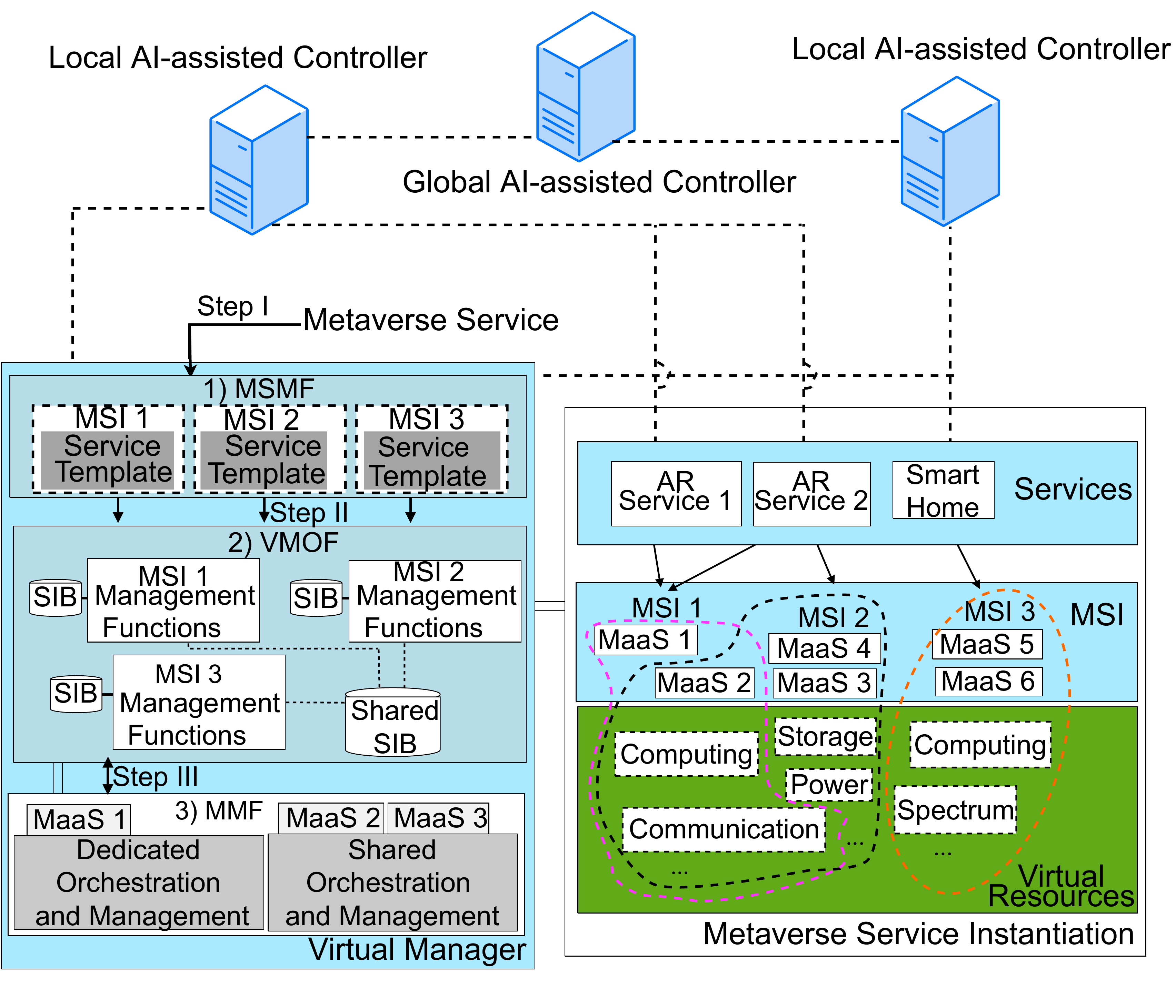} 
\caption{Instantiation process of Metaverse service under Slicing4Meta architecture.}
% \label{fig:system model}
\vspace*{-4mm}
\end{figure}

\subsection{Intelligent MSIs Integration of Slicing4Meta}

% As shown in Fig. 3, AI-assisted controllers mainly play important roles in the phases of preparation (Step I), planning (Step II), and run-time (Step III) during the life-cycle of MSIs 
As shown in Fig. 3, the life-cycle of MSIs includes the phases of preparation (Step I), planning (Step II),
and run-time (Step III) \cite{wu2022ai}. Specifically, in the preparation phase, the global AI-assisted controller predicts the service type and the QoE requirements of Metaverse users via using prediction technologies (\emph{e.g.}, data mining). Specifically, the global AI-assisted controller first integrates and designs the necessary Metaverse environment. Then, the global AI-assisted controller sends the design results to the associated local AI-assisted controller. In the planning phase that involves the real-time instantiation, configuration and activation of MSIs, the local AI-assisted controller creates and configures MSIs by dynamically creating, revising and adjusting dedicated/shared MaaS models and the corresponding resources via online AI tools. In the run-time phase, MSIs are capable of handling service flows to support Metaverse services of certain types, where the local AI-assisted controller is responsible for supervision (\emph{e.g.}, monitoring service flow and resource utilization) and modification (\emph{e.g.}, scaling MSI cluster). Meanwhile, the local AI-assisted controller timely reports the supervision and modification results to the global AI-assisted controller and obtains the feedback. Note that the continuous re-assessment of service deployment and resource utilization can take place in the same MSI cluster through the fine-grained approaches in a short timescale. While the reservation of aaS models, corresponding resources and technologies are performed in a long timescale through the coarse-grained approaches according to the requirements from Metaverse service clusters.

\section{Two Typical Types of Metaverse Services}
In this section, we elaborate how to fulfill the requirements of the two major types of Metaverse services under the Slicing4Meta architecture, as shown in Table I.
% , the QoE requirements, involved heterogeneous resources, related MaaS models, and the corresponding Metaverse scenarios are listed.
\begin{table*}[htbp] \label{tab:notations}
\centering
\vspace{-5mm}
\caption{Metaverse Scenarios, QoE requirements, and the corresponding MaaS models and multi-dimensional resources}
\vspace{-2mm}
\setlength{\tabcolsep}{1mm}{
\begin{tabular}{cl}%{p{1.5cm}p{7cm}}
\centering
\includegraphics[width=13cm]{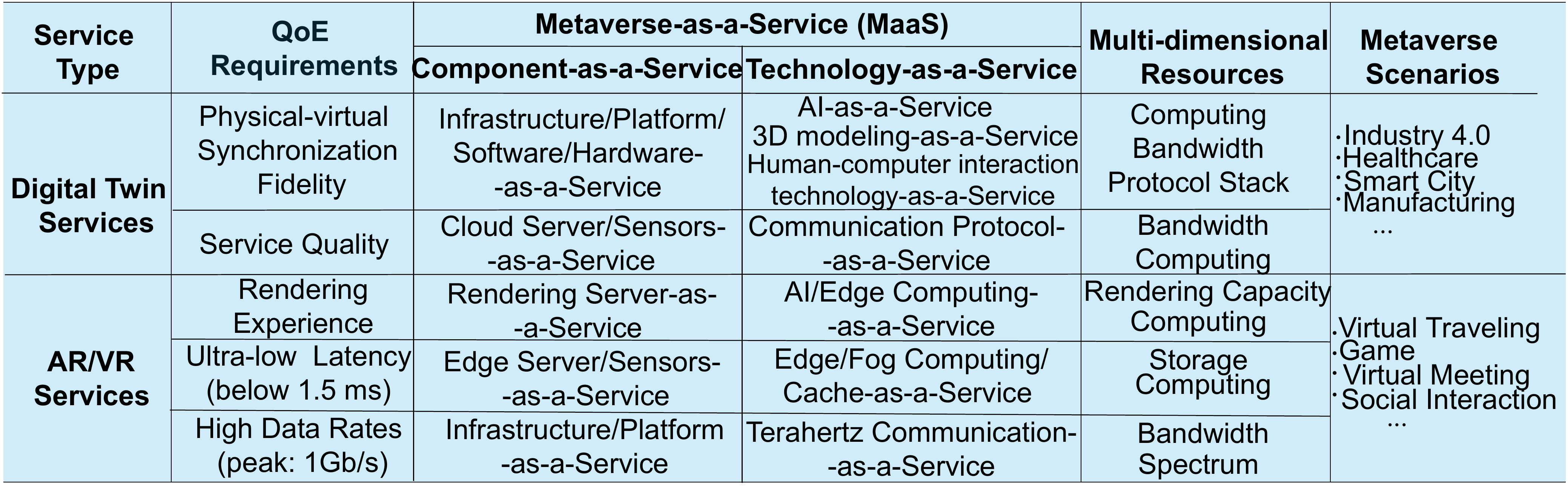} 
\vspace{-5mm}
\end{tabular}
}
\end{table*}

\subsection{Digital Twin Services}
% Digital twin (DT), a virtual representation of a physical object, is intensely spawning DT services which could serve any phase of product development such as simulating tests on virtual prototypes and predicting the actual performance of the physical objects. 
Many highly dynamic and complex scenarios such as Industry 4.0 and manufacturing, require new service dimensions and experiences. Digital twin (DT), a virtual representation of a physical object, is promising to cope with the aforementioned trends by running and simulating virtual replicas before physical objects are built/deployed. Naturally, DT services become an essential booster for facilitating DT development. Specifically, DT services can serve any phase of a physical object's life-cycle such as design, simulation, monitoring and running. Meanwhile, a large number of DT services, such as knowledge (part of data that may be extracted from existing data) services, are needed to build a functioning DT virtual world \cite{QI20213}. It is thus important to customize specific DT services by integrating and customizing various DT MSIs to guarantee physical-virtual synchronization fidelity. As shown in Table \uppercase\expandafter{\romannumeral1}, necessary MaaS models for DT services are given from the following two aspects:

% Many highly dynamic and complex scenarios such as Industry 4.0, manufacturing,  and smart city, require new service dimensions and QoE such as modeling, monitoring, and even improving data-driven decision-making processes
% supporting DT services mainly involve integrating, creating and building
% including technologies, components, and even knowledge, which could be packaged and delivered as a service model. 
% The required MaaS models are classified into the following three aspects:

\begin{enumerate}
\def\labelenumi{(\arabic{enumi})}
\item
\emph{Component-as-a-service}: 
% As the key basis to support DT services, both the physical and virtual components provide specific capabilities (\emph{e.g.}, computing, transmitting and analysis) to guarantee the QoE of users. 
Many components in DT systems, such as hardware (\emph{e.g.}, sensors and graphics processing units), software and infrastructures (\emph{e.g.}, virtual machines), can be regarded as service models to provide specific capabilities (\emph{e.g.}, computing and analysis) by providing needed multi-dimensional resources. For example, sensor-as-a-service (\emph{e.g.}, Nexxiot's sensors) provides specific communication protocols to support continuous data collection on physical objects.

% To facilitate the integration of processing capabilities and the integration of resources

% could be regarded as service models.

% Some resource perception and assessing technologies such as sensor and adapters could  resource packaging technologies such as web services, 
\item
\emph{Technology-as-a-service (TaaS)}: Some TaaS allows users to customize on-demand DT services by consuming multi-dimensional resources, where TaaS mainly performs the following functions: 1) Connecting the physical-virtual world and various DT systems, such as human-computer interaction technology-as-a-service. 2) Managing massive data (collection, fusion, processing and analyzing) such as data transmission protocol-as-a-service. In addition, some TaaS such as time division and frequency division, can also be delivered as service models.

% To connect the physical and virtual world, some 

% \item
% \emph{Data as a service}:

\end{enumerate}

\subsection{AR/VR Services}
% AR/VR services offer Metaverse users immersive experiences to visualize actionable insights, where both the user devices and AR/VR services should be popularized and functionalized to derive evolution to the immersive phases and guarantee QoE requirements 

AR/VR services offer Metaverse users immersive experiences, where both the AR/VR devices and services should be popularized and functionalized to derive evolution to the immersive phases \cite{dionisio20133d}. Many AR/VR devices in the PWL layer are licensed on a specific basis to deliver user-related features (\emph{e.g.}, user feelings and interaction signals), which require real-time interactive experiences \cite{8869705,speculative,9382026}. 
% in the MSIL, MSIs customized for specific AR/VR services, pose stringent QoE requirements (\emph{e.g.}, peak data rates up to 1 Tb/s and user-experienced data rates up to 1 Gb/s).
Moreover, in the MSIL, specific AR/VR services pose stringent QoE requirements (\emph{e.g.}, peak data rates up to 1 Tb/s) \cite{8869705,speculative,9382026}. To cope with these requirements, there is a significant need for integrating MaaS models and the associated multi-dimensional resources. By sharing and/or purchasing the integrated MaaS models on demand, businesses and/or operators can reduce CAPEX/OPEX costs while meeting the QoE requirements. Under this circumstance, we classify the necessary MaaS models to support customized AR/VR services as follows:

% Meeting these requirements boots a significant need for integrating MaaS models and the associated multi-dimensional resources. 

\begin{enumerate}
\def\labelenumi{(\arabic{enumi})}
\item
\emph{CaaS}: First, AR/VR devices are the basic CaaS models to support feature tracking. Second, similar to DT services, software, platforms, infrastructures and hardware can be integrated as services to optimize user engagement. For example, AR visualization software can be integrated with AR Software Developer Kit (ADK) platform to create a personalized MSI model to meet the specific AR interaction experience.

\item
\emph{TaaS}: AR/VR scenarios such as video shooting and social interaction, require highly dynamic and complex technologies to support latency-sensitive AR/VR services. Therefore, technologies such as content creation technology (\emph{e.g.}, three-dimensional modeling and holography communication) and rendering processing technology (\emph{e.g.}, cloud rendering) can be delivered and/or integrated as service models \cite{dionisio20133d}.
% to meet specific QoE requirements (\emph{e.g.}, latency and data rates) of AR/VR services . 
\end{enumerate}

% 

% can determine user motions and send interaction signals to AI-assisted controllers. 

%  by a wearable or a hand-held AR/VR devices

\section{Case Study For Slicing4Meta}
% By integrating, creating and modifying the afore-mentioned CaaS models and TaaS models, the global AI-assisted controller cooperates with local AI-assisted controllers to make tailoring decisions on deploying MSIs to support specific Metaverse services.
% To demonstrate the effectiveness of our Slicing4Meta architecture, we present a virtual travel case study and corresponding simulation results in this section. 

\subsection{Special Case}
% Under the Slicing4Meta architecture, SPs purchase MaaS models from businesses and/or operators to construct specific MSIs to support Metaverse services. Therefore,
% Constructing specific MSIs involves fully utilizing MaaS models, allocating optimally multi-dimensional resources, and guaranteeing QoE experiences of users. Here 
We consider a basic virtual travel case on a beach, including two main Metaverse services (\emph{i.e.}, AR service and virtual-based service) \cite{um2022travel}. Specifically, the AR service provides AR navigation services, AR maps, and extended reality experiences. Virtual-based services provide ``beach experiences" by exploring a virtual world freely as avatars, which enables users to create and experience a virtual beach without limits in the virtual world and deliver the experiences back.

% We consider a basic AR/VR service that provides users virtual immersion experiences via virtual travel \cite{um2022travel}. In the following, we list necessary resources (also called capabilities) and the associated MaaS models that are needed to construct the travel MSI:

$\bullet$ Computation, storage and communication resources. Guaranteeing fluid synchronization between Metaverse users’ movements and visual perception is of importance for virtual travel. The synchronization includes complex procedures such as predicting user behavior and processing user interaction, which requires a large amount of computation, storage and network resources. Generally, IaaS and PaaS are expected to provide these required resources which are first integrated into a virtual resource pool. Then, the global AI-assisted controller cooperates with local AI-assisted controllers to make decisions on deploying IaaS/PaaS models and the needed resources for virtual travel MSIs.

% constituting the resources into various IaaS and PaaS models
 % and then constituted into various IaaS and PaaS models. Then  allocating these models to various virtual travel MSIs 
% deploying various virtual travel MSIs according to the QoE requirements of users such as tolerable ambient clarity and latency. 

% involves processing phases such as predicting, caching and analysis. 

% Many tasks (\emph{e.g.}, predicting, processing, caching and load balancing) require 

% Virtual travel requires 
% To provide engaging travel content and support your journey to travel 
% The platform allows you to plot your journey as you explore five key locations across the reef, provides information about each area and how coral and certain species live, as well as giving users the option to listen into the sounds of the ocean
% Alternatively, check out the Ocean Agency’s 360-degree images on Google Earth and AirPano, which allows you to ‘swim’ through a reef in Indonesia using interactive images.

% By integrating 

% to form specific MSIs in the virtual travelling scenarios

Data processing capabilities. (1) Data collection: The physical-virtual connection empowers the data collection process to update virtual representations (\emph{e.g.}, travel conditions). Specifically, from the physical world, data such as the visual and audio of users and the service requirements can be collected by AR/VR devices. From the virtual world, data such as weather and views can be monitored and simulated by virtual models. Therefore, device-as-a-service is needed to create/modify device patterns to provide specific data collection capabilities and interaction protocols, while monitoring-as-a-service is needed to monitor data and the corresponding functionalities. (2) Data fusion: The data collected from different sources (\emph{e.g.}, physical world and virtual world) can be merged via data fusion algorithms to make the data more informative. Therefore, data management-as-a-service is needed to provide centralized storage for various data sources. 
% Additionally, AI algorithms such as neural networks and clustering algorithms can be used as services to facilitate data fusion and monitoring. 
Meanwhile, AI technologies such as machine learning can be incorporated into controllers to realize intelligent data analysis, fusion and management.

Undoubtedly, during the construction process of the travel MSIs, some MaaS models supply multi-dimensional resources while others consume these resources. Therefore, before constructing specific MSIs via MaaS models, we should quantitatively analyze the QoE when integrating multi-dimensional resources of various MaaS models. In this paper, similar to the idea of \cite{du2021optimal}, we exploit a novel metric that consists of subjective experience and objective service quality, named Meta-Immersion (MI), to model the QoE of Metaverse users. 

\begin{figure}[htbp]
\centering
\vspace*{-2mm}
\includegraphics[width=7cm]{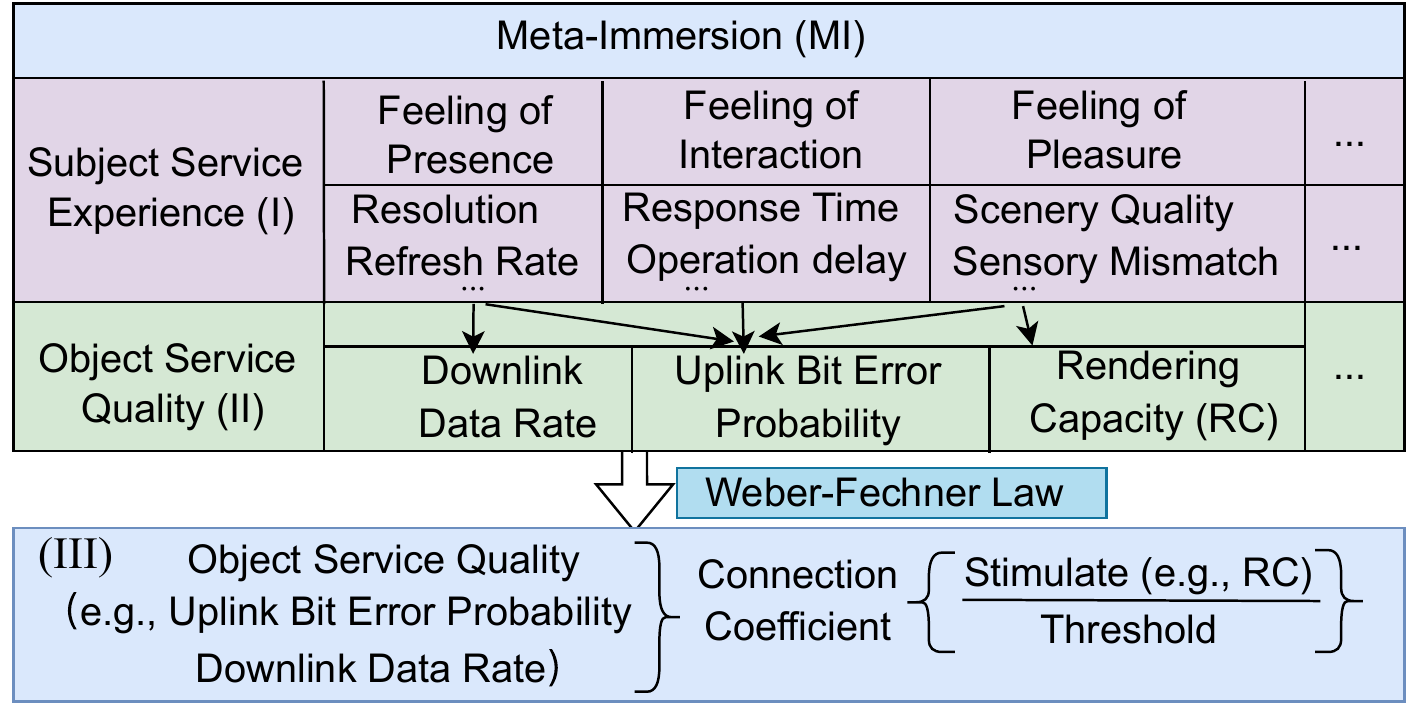} 
\caption{A novel QoE metric in Metaverse: Meta-Immersion, and corresponding experience factors, service indicators and performance measures.}
\label{fig:system model}
\vspace*{-2mm}
\end{figure}

As shown in Fig. 4, for the QoE of virtual travel services, the objective service quality is measured by the downlink rate, uplink bit error probability (BEP) and rendering capacity (Part \uppercase\expandafter{\romannumeral2}).
% , which impact the interaction (\emph{e.g.}, response time and refresh rate) between the virtual and physical worlds. 
Meanwhile, the subjective experience is affected by the feelings of travelers in terms of presence, interaction and pleasure, which can be reflected in the object service quality (Part \uppercase\expandafter{\romannumeral1}). For example, if a traveler is visiting a virtual beach, the traveler can feel the wind and sea waves, where the visual, auditory, and touch experiences of travelers can be affected by the rendering capacity. To derive the QoE in virtual travel, we establish the relationship between the objective service quality and the subjective experience (Part \uppercase\expandafter{\romannumeral3}). Specifically, we use Weber–Fechner law to express the relationship between the stimulus of rendering experiences that travelers perceive from virtual travel and the perceived subjective feelings within the human sensory system. Formally, the difference perception is directly proportional to the relative change of the rendering stimulus \cite{du2021optimal}. Moreover, we use a normalization function to express the objective stimulus from the physical world. As shown in Part III, by combining the rendering and objective stimulus, we derive the expression of MI. When the rendering stimulus and/or objective stimulus change by more than a certain proportion of its actual magnitude, the AI-assisted controllers adjust the interaction and the objective service quality by integrating MaaS models in both virtual and physical worlds. 

\subsection{Numerical Results}
% In this simulation, we evaluate the MI of 30 users travel in the virtual world by AI-assisted controllers integrating rendering capacity from rendering servers and bandwidth on the downlink from infrastructures. 

%  The local AI-assisted controllers (\emph{e.g.}, Sysmac AI-assisted controllers) are used to fuse and analyze the data collected from different sources (\emph{e.g.}, physical world, virtual world and travel QoE requirements) via AI technologies (\emph{e.g.}, deep learning and computer version). By fusing and analyzing the data fused by the local AI-assisted controllers, the global AI-assisted controller with machine learning modules makes the global decision for uniformly allocating rendering capacity to the specific travel MSIs. Once receiving the global decision, the local AI-assisted controllers allocate rendering capacity to the corresponding IaaS (\emph{e.g.}, rendering servers and base stations) via the Uniform allocation scheme.

% We evaluate the MI of 30 users travel in the virtual world, with the aim to maximize total QoE of these users by creating specific travel MSIs. In the simulation, 
By evenly allocating rendering capacity to users, we examine the MI under four different data rate conditions with respect to the number of users via simulation experiments. 
% In this simulation, we uniformly allocate rendering capacity to users. 
% , where the data rates are fixed
We consider the total rendering capacity of the rendering server is set to 4000 K, where 1 K resolution refers to 960 $\times$ 480 pixel resolution. Moreover, there are 56 virtual objects in the virtual travel scenario, where the number of virtual objects for a user is randomly chosen from $\{1, 2,\ldots, 55, 56\}$ and the rendering capacity of each virtual object is set to 20K. Fig. 5 shows the MI (QoE) varying with the number of users under four different data rate conditions. As shown in Fig. 5, we find that when the total amount of rendering capacity is fixed (4000 K), Metaverse users with higher data rates always obtain better MI. In other words, when certain resources of MaaS models are insufficient, we can increase the resources by modifying the MaaS models or supplement the consumption from other MaaS models at the cost of other-dimensional resources.

\begin{figure}[htbp]
\centering
\vspace*{-4mm}
\includegraphics[width=6cm]{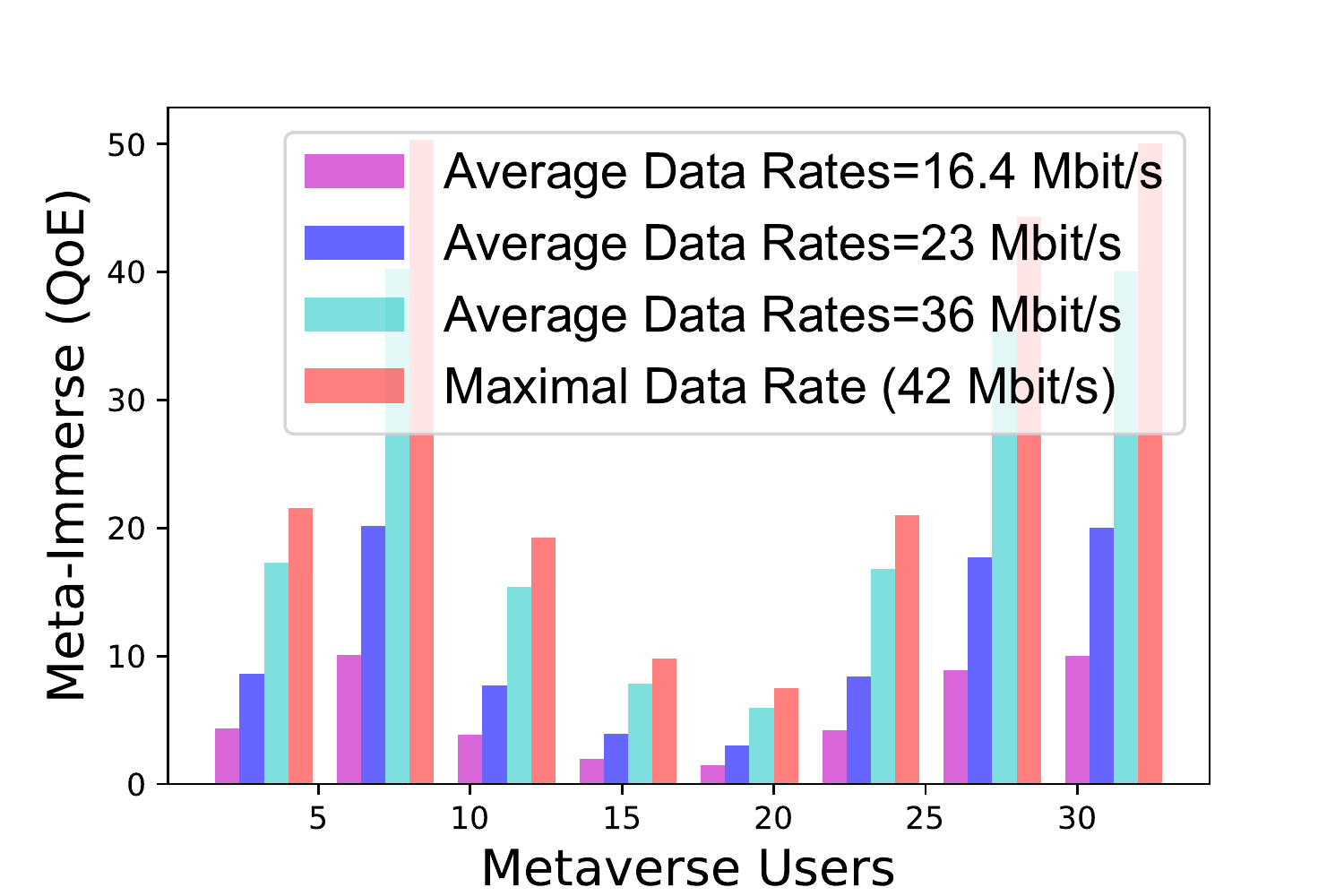} 
\caption{Meta-Immerse under four different data rate constraints.}
\label{fig:system model}
\vspace*{-5mm}
\end{figure}

% there exists a trade-off between rendering capacity and bandwidth on the downlink. 

%  can be supplemented 
% the Integration scheme always outperforms the Uniform scheme in terms of the total MI when the total amount of rendering capacity and bandwidth are fixed. 

\section{Open Challenges and Solutions}
Although the aforementioned good aspects of integrating MaaS models under the Slicing4Meta architecture can be expected for supporting the potential Metaverse services, it also brings some challenges.
% , such as isolation and security. 
% In this section, we present a discussion of isolation and security challenges and some potential solutions to facilitate the implementation of Slicing4Meta for the evolution to 6G.

\subsection{Isolation}
An MSI customized for a specific service should operate independently from other MSIs. Therefore, 
% to guarantee the independent operation of all MSIs, 
isolation should be guaranteed for diversified Metaverse services. Specifically, as various MSIs are built upon a common shared physical network, physical isolation (\emph{e.g.}, isolate physical resources) and logical isolation (\emph{e.g.}, split time and frequency) are always needed to avoid signal interference and resource conflicts. Moreover, as various MSIs may share the same MaaS models and virtual resources, scheduling isolation (\emph{e.g.}, resource scheduling) is important at the orchestration and management level, where policy-based orchestration mechanisms should be developed. Additionally, a certain isolation degree for the type of Metaverse services is also needed, where the lower the isolation degree, the easier to share MSIs and MaaS models.

\subsection{Security and Privacy}
% Under Slicing4Meta, security and privacy are always imperative issues. 
As the Slicing4Meta architecture is based on mobile wireless networks, an attack (\emph{e.g.}, single point of failure and distributed denial-of-service attacks) is a threat to both physical and virtual worlds. Moreover, as the massive amounts of data generated in the virtual and physical worlds involve important and sensitive information, the confidentiality, integrity, and availability of data should be guaranteed. When the privacy data of user behavior is frequently transformed into digital assets, data management and protection should be enhanced.

Essentially, a few emerging technologies can be configured into MSIs as a service. For example, blockchain can store submitted transactions to enable tracking and protection of digital assets. Moreover, it can store digital biometrics-based identities as well as physically authenticate identities  \cite{Minrui}.
% Furthermore, blockchain can be regarded as a complete economic service to connect the virtual world and real world, where users are allowed to trade virtual items in the same way as in the real world \cite{Minrui}. 
Zero trust eliminates the theft of sensitive information through continuous authentication and verification.

\section{Conclusion}
In this paper, we have proposed MaaS to provide on-demand subscription solutions for customizing Metaverse services. To unify the orchestration and management of MaaS models, we have proposed a Slicing4Meta architecture, in which two-tier AI-assisted controllers are used to facilitate the judicious and timely coordination of MSIs. Under the Slicing4Meta architecture, we have specified two typical types of Metaverse services based on various QoE requirements and presented the specific MaaS models needed for each type of services. Moreover, we have illustrated a virtual travel case, where we quantitatively examine the relationship between the QoE of users and the multi-dimensional resources via simulation results. Finally, we have discussed the isolation and security/privacy issues of Slicing4Meta and proposed potential solutions to address these issues.

\bibliographystyle{IEEEtran}
\bibliography{reference}

\begin{IEEEbiographynophoto}
{Yi-Jing Liu} is a Research Collaborator with the School of Computer Science and Engineering, Nanyang Technological University, Singapore. She is pursuing a Ph.D degree at the National Key Laboratory of Science and Technology on Communications, University of Electronic Science and Technology of China, China.\\
\\\textbf{Hongyang Du}
is currently pursing the Ph.D degree at the School of Computer Science and Engineering, Nanyang Technological University, Singapore. \\
\\\textbf{Dusit Niyato}
is a President’s Chair Professor with the School of Computer Science and Engineering, Nanyang Technological University, Singapore. He is an IEEE Fellow. \\
\\\textbf{Gang Feng}
is a Professor with the National Laboratory of Communications, University of Electronic Science and Technology of China, China.\\
\\\textbf{Jiawen Kang}
is a Professor with the School of Automation, Guangdong University of Technology, China. \\
\\\textbf{Zehui Xiong}
is an Assistant Professor with the Pillar of Information Systems Technology and Design, Singapore University of Technology and Design, Singapore.

\end{IEEEbiographynophoto}

% Y. J. Liu 

% H. Du and D. Niyato are with School of Computer Science and Engineering, Nanyang Technological University, Singapore. Y. J. Liu is also with the National Key Lab on Communications, University of Electronic Science and Technology of China, China. Gang Feng is with the National Key Lab on Communications, University of Electronic Science and Technology of China, and also with the Yangtze Delta Region Institute (Huzhou), University of Electronic Science and Technology of China, China. J. Kang is with the School of Automation, Guangdong University of Technology, China. Z. Xiong is with the Pillar of Information Systems Technology and Design, Singapore University of Technology and Design, Singapore. G. Feng is the corresponding author (e-mail: fenggang@uestc.edu.cn).

% \begin{IEEEbiographynophoto}
% {Authors' Bio}

% \end{IEEEbiographynophoto}

%

\end{document}